\documentstyle[12pt]{article}
\def\beqa{\begin{eqnarray}}
\def\eeqa{\end{eqnarray}}
\def\beq{\begin{equation}}
\def\eeq{\end{equation}}
\begin{document}
\renewcommand{\theequation}{\thesection.\arabic{equation}}
\begin{titlepage}
        \title{Nonminimal Derivative Couplings and Inflation in
              Generalized Theories of Gravity}
\author{S. Capozziello$^{1}$\thanks{E-mail: capozziello@vaxsa.csied.unisa.it},
G. Lambiase$^{1}$\thanks{E-mail: lambiase@physics.unisa.it}~~and H.-J.
Schmidt$^{2}$\thanks{E-mail: hjschmi@rz.uni-potsdam.de}
\\ {\em $^1${\small Dipartimento di Scienze Fisiche ``E.R. Caianiello"}} \\
 {\em{\small Universit\'a di Salerno, 84081 Baronissi (Sa), Italy.}} \\
 {\em{\small Istituto Nazionale di Fisica Nucleare, Sez. di Napoli, Italy.}} \\
 {\em $^2${\small Institut f\"ur Mathematik, Universit\"at Potsdam}}\\
{\em{\small PF 601553, D-14415 Potsdam, Germany}}\\
 {\em{\small Institut f\"ur Theoretische Physik, Freie Universit\"at Berlin,}}\\
{\em{\small Arnimallee 14, D-14195 Berlin, Germany}}\\ }
\date{\today}
\maketitle
\begin{abstract}
We study extended theories of gravity where nonminimal derivative couplings of the form
$R^{kl}\phi_{,\,k}\phi_{,\,l}$ are present in the Lagrangian. We show how and why the
other couplings of similar structure may be ruled out and then deduce the field
equations and the related cosmological models. Finally, we get inflationary solutions
which do follow neither from any effective scalar field potential nor from a
cosmological constant introduced ``by hand'', and we show the de Sitter space--time to
be an attractor solution.
 \end{abstract}
\thispagestyle{empty} \vspace{20.mm} PACS number(s): 04.50.+h, 98.80.Cq\\
 \vspace{5.mm} Keyword(s): Cosmology, Alternative Theories of Gravity,
  Nonminimal Coupling.
 \vfill
\end{titlepage}

\section{Introduction}
\setcounter{equation}{0} The existence of an inflationary phase in the early universe is
now generally accepted, cf. e.g. \cite{Li}. The first idea how to generate an
inflationary de Sitter phase was to  introduce a cosmological term $\Lambda>0$. However,
this idea suffered from the cosmological constant problem: In natural units, one gets
$\Lambda\simeq 10^{-128}$, an almost unexplainable fine--tuning would be necessary to
achieve that.

One of the next ideas to get an inflationary phase was to introduce other fields or
other gravitational field equations. For instance, the Starobinsky model has a
gravitational Lagrangian\footnote{We choose sign conventions such that the de Sitter
spacetime has a curvature scalar $R<0$, and the $+$-sign in Eq.(\ref{1}) shows that we
restrict to the tachyonic--free case.}
 \beq \label{1}
 L=-\frac{R}{2}+\frac{l^2}{12}R^2\,,
 \eeq
 where $l$ is a constant of length--dimension and we have no cosmological constant.

In \cite{MS} it was shown, that, nevertheless, the de Sitter spacetime with
the effective value of $\Lambda$ depending on $l$ is a transient attractor
of the corresponding fourth--order vacuum field equations, cf.\cite{Sch}
for further details and references.

Soon it became clear how this behavior can be explained: By a conformal transformation
(see e.g.\cite{St}\footnote{This conformal relation was independently found by several
authors; it should be called Bicknell-theorem.}) which has an almost constant conformal
factor near the de Sitter spacetime, the models with $L=f(R)$ (where $f$ is nonlinear in
$R$), can be transformed to Einstein's theory with a minimally coupled scalar field
$\phi$ and a potential $V(\phi)$ describing its self--interaction. In regions of an
almost constant positive potential $V$, we can interpret $V(\phi)$ as an effective
cosmological constant leading to a quasi de Sitter phase.

On another branch of research, also nonminimally coupled scalar fields have been used to
deduce the inflationary phase, i.e.
 \beq \label{2}
 L=F(\phi, R),\quad \mbox{with} \quad
 \frac{\partial^2 F}{\partial \phi\partial R}\neq 0\,,
 \eeq
 i.e., this Lagrangian does not have the form of ${\displaystyle L=f(R)+V(\phi)}$.
However, also this kind of theories is conformally related (up to singular exceptions)
to the theories mentioned above, but they deserve a lot of consideration since they
allow, several times,  to get inflation without the ``graceful exit'' problem bypassing
the shortcomings of former inflationary models (see e.g. \cite{LA} for extended and
hyperextended inflation).

However, the form of nonminimal coupling besides higher--order terms in the effective
gravitational Lagrangian can be chosen in several ways (see e.g. \cite{ACLO}) to obtain
one or more than one inflationary phases but the ingredients, after a conformal
transformation are always the same: Inflation is driven by a scalar field potential
which, for a certain period, assumes the appearance of an effective  cosmological
constant.

Our issue is now: Is it possible to recover the cosmological constant, and then the
inflationary phase ``without'' considering any effective potential? In \cite{CD} and
\cite{CDM}, it was discussed how to construct an effective cosmological constant
starting from extended gravity theories (i.e. nonminimally coupled or higher--order
theories). As the main result, an extension of the cosmic no hair conjecture was found.
In any case, the scalar field potential in nonminimally coupled theories or the
conformally related scalar field potential for higher--order theories were essential
features. In \cite{CD}, it was shown that for nonminimally coupled theories without a
scalar field potential (e.g. a pure Brans--Dicke theory) an effective cosmological
constant is never recovered.

In spite of this result, in \cite{CL}, it was shown that an effective cosmological
constant can be recovered if a nonminimal derivative coupling is introduced in the
gravitational Lagrangian also if no scalar field potential  or higher--order terms in
curvature invariants are taken into account. In other words, it seems that a new type of
inflation can be dynamically induced just by considering the self--coupling between
geometry and the kinetic term of some given scalar field.

 In 1993, Amendola \cite{A} started to consider further types
of coupling between curvature and the scalar field, called nonminimal derivative
coupling, see also \cite{CL} for details. The main ingredient of this kind of couplings,
already mentioned in \cite{Li}, Eq.(9.5.9), reads
 \beq \label{3}
 L_1=R^{kl}\phi_{,\, k}\phi_{,\, l}\,.
 \eeq

The aim of the present paper is to study this kind of couplings and connect them with
inflation.

The paper is organized as follows: In Sect. 2, we consider all Lagrangians of type
(\ref{3}) and find out which of them are really independent. In Sect. 3, the field
equations are deduced; Sect. 4 deals with the corresponding cosmological models.
Discussion and conclusions are drawn in Sect. 5. In particular, we discuss the results
in relation to  the  transformations \beq \label{4} \tilde{g}_{ij}=\frac{\partial {\cal
L}}{\partial R_{ij}}\,, \eeq see \cite{Sok} and \beq \label{5}
\hat{g}_{ij}=g_{ij}+\lambda^{2}u_{i}u_{j}\,. \eeq see \cite{CLS} asking for a
generalization of conformal transformations in which it is possible to find out the
analogous ingredients of a scalar field potential.

\section{ The Possible Lagrangians}
\setcounter{equation}{0}

The following six terms carry a geometric structure similar to $L_1$, Eq. (\ref{3}):
 \beqa
 L_2&=&R\phi_{,\,k}\phi^{,\,k} \label{6} \,{,}\\
 L_3&=&R\phi\Box \phi \label{7} \, , \\
 L_4&=&R^2\phi^2\label{8}\,, \\
 L_5&=&R_{,\,k}\phi^{,\,k}\phi\label{9}\,, \\
 L_6&=&R^{kl}\phi\phi_{;\,kl}\label{10}\,, \\
 L_7&=&\phi^2\Box R\label{11}\,.
 \eeqa
 This list is complete in the sense that every scalar of the same geometric structure
can be written as linear combination of $L_1, \ldots, L_7$\footnote{This list is almost
identical to the list Eq. (\ref{2}) of Ref. \cite{A}; the difference is the following:
our $L_4=R^2\phi^2$ is absent in \cite{A}, because there the appearance of derivatives
was required, whereas we apply  the more geometric point of view that $R^2$ and $\Box R$
have the same geometric structure, and so $L_4$ has to be included if we have
$L_7=\phi^2\Box R$. The fact that 3 of these terms may be neglected due to divergences,
was deduced on another way in \cite{A}. }. To find out an independent subset of $L_1,
\ldots, L_7$, we apply the fact that the addition of a divergence does not alter the
field equation and is therefore not necessary for our purposes.

Using the divergencies
 $$
 (R\phi^{,\, k}\phi)_{;\, k}\,,\quad (R^{ik}\phi\phi_{,\, k})_{;\, i}\,,
 \quad (R^{,\, i}\phi^2)_{;\, i}\,{,}
 $$
 we conclude that without loss of generality, $L_5$, $L_6$ and $L_7$
are not necessary to be considered. Though $\Box (R\phi^2)$ represents a divergence of
the same structure, it does not further reduce the necessary set $\{L_1, \ldots, L_4\}$.
$L_4$ may be ruled out because it has already the structure of Eq. (\ref{2}). Further, $
L_3$ is only marginally interesting here, because it contains also $\phi$ itself, and we
are mainly interested in a coupling, where only the gradient of $\phi$ is included.

Therefore, our main topic is to consider ${\cal L}_1$ and ${\cal L}_2$ (and sometime
${\cal L}_3$, too). That these three Lagrangians are really independent will become
clear after having deduced the field equations.

\section{ How to Deduce the Field Equations}
\setcounter{equation}{0}

In subsection 3.1 we apply the variational derivative $\delta/\delta\phi$ and in 3.2
analogously $\delta/\delta g_{ij}$ to the Lagrangian density ${\cal L}_i=\sqrt{-g} L_i$,
$i=1,2,3$, cf. Eqs. (\ref{1}), (\ref{6}), (\ref{7}).

\subsection{ Field Equation for the Scalar Field}

From ${\cal L}_3$ we get
 \beq\label{12}
 0=2R\Box \phi+\phi\Box R+2R_{,\,k}\phi^{,\, k}
 \eeq
 which can be written in a more compact form as
 $$
 0=R\Box \phi+\Box (R\phi).
 $$
 For ${\cal L}_1$ and ${\cal L}_2$ we give a common deduction. To this end we define the
tensor
 \beq\label{13}
 V^{kl}=R^{kl}+\alpha Rg^{kl}\,{,}
 \eeq
 where $\alpha$ is a constant, and $L_0=L_1+\alpha L_2$, i.e.
 \beq\label{14}
 L_0=V^{kl}\phi_{,\, k}\phi_{,\, l}\,{,}\quad
 {\cal L}_0=\sqrt{-g}L_0\,{.}
 \eeq
 Using the formula
 $$
 \frac{\partial {\cal L}_0}{\partial \phi_{,\, i}}=2\sqrt{-g}V^{ik}\phi_{,\, k}\,{,}
 $$
 we get the result that $0=\delta {\cal L}_0/\delta\phi$ then reads
 \beq\label{15}
 0=-2(V^{ik}\phi_{,\, k})_{;\, i}\,{.}
 \eeq
 From the contracted Bianchi identity, we get
 $$
 V^{ik}_{\phantom{ik};\, k}=\left( \frac{1}{2}+\alpha\right)R^{;\, i}
 $$
 so $\alpha =-1/2$ plays a special role\footnote{This is the same case in Ref. [6, eq.
(26)]}. The field equation (\ref{15}) for $\phi$ can be rewritten as (after dividing by
-2)
 \beq\label{16}
 0=R^{ik}\phi_{;\, ik}+\alpha R\Box \phi+\left(\frac{1}{2}+\alpha\right)R^{,\,
 k}\phi_{,\, k}\,{.}
 \eeq

\subsection{The Gravitational Field Equation}

Let us start with the easier-to-deal case $L_2=R\psi$, with
 \beq\label{17}
 \psi =\phi_{,\, k}\phi_{,\, l}g^{kl}\,{.}
 \eeq
 The field equation shall be deduced in 2 steps: First we consider the intermediately
introduced auxiliary field $\psi$ as an independent scalar field
(and that problem is of the known structure Eq. (\ref{2})), and
second, we add the correction term $-R\phi^{,\, a}\phi^{,\, b}$
which results from the fact that $\psi$, Eq. (\ref{17}), {\it has}
a dependence on $g^{kl}$. As a result we get\footnote{Eq.
(\ref{18}) is identical to Eq. (\ref{15}) of Ref. \cite{A}, which
was given there without detailed explanation. Eq. (\ref{19}),
however, we did not find in the literature. }
 \beq\label{18}
 E_2^{ab}=\frac{1}{2}g^{ab}R\phi_{,\, k}\phi^{,\, k}-R\phi^{,\, a}\phi^{,\, b}-
 R^{ab}\phi_{,\, k}\phi^{,\, k}+(\phi_{,\, k}\phi^{,\, k})^{;\,ab}-g^{ab}\Box
 (\phi_{,\, k}\phi^{,\, k})\,{,}
 \eeq
 where
 $$
 E_i^{ab}=\frac{1}{\sqrt{-g}}\frac{\delta {\cal L}_i}{\delta g_{ab}}\,{.}
 $$
 The same principle applies to $L_1=R_{kl}u^{kl}$. First, $u^{kl}$
is considered as any contravariant symmetric tensor field, and
second, the fact that $u^{kl}=\phi^{;\, k} \phi^{;\, l}$ depends
on the metric (because the dependence of the full Lagrangian is on
$\phi_{,\, k}$ and not on $\phi^{,\, k}$) via
 $$
 u^{kl}=g^{ka}\phi_{;\, a}g^{lb}\phi_{;\, b}\,{,}
 $$
 one has the correction terms $-R^{ak}\phi_{;\, k}\phi^{;\, b}-
R^{bk}\phi_{;\, k}\phi^{;\, a}$. The final expression is this one
 \beqa
 E^{ab}_1&=&\frac{1}{2}g^{ab}R^{kl}\phi_{, \, k}\phi_{,\, l}-\frac{1}{2}
 \Box(\phi^{,\, a}\phi^{,\, b})-\frac{1}{2}g^{ab}(\phi^{,\, k}\phi^{,\,l})_{;\, kl}\,\label{19} \\
         & &-[(\phi^{,\, k} \phi^{,\, (a})^{;\, b)}]_{;\,k}-2R^{k(a}\phi_{,\, k}\phi^{,\, b)}\,,\nonumber
 \eeqa
 where round brackets  denote symmetrization. It
should be noted that Eq. (\ref{19}) has already a quite compact form. After multiplying
out the derivatives one gets much more terms, and if one changes the ordering of the
covariant derivatives one would  produce extra terms like
 $$
 \phi^{,\, c} \phi^{,\, d}R^{a\phantom{c} b}_{\phantom{a} c
 \phantom{b} d}\,{.}
 $$
 Both for Eqs. (\ref{18}) and (\ref{19}) the highest
$\phi$--derivative is a third one.

\section{Cosmological Models}
\setcounter{equation}{0}

In \cite{CL} the following Lagrangian ${\cal L}=\sqrt{-g}L$ has
been discussed, where
 \beq\label{20}
 L=-\frac{R}{2}+\frac{1}{2}g^{kl} \phi_{,\, k} \phi_{,\, l} +
 \zeta L_1+\xi L_2\,{,}
 \eeq
 where $L_1$ and $L_2$ are defined by Eqs. (\ref{3}) and (\ref{6}).
From dimensional reason, $\zeta$ and $\xi$ have the dimension $l^2$, where $l$ is any
length, $L_1$, $L_2$ represent corrections to the Lagrangian of Einstein's theory
without the $\Lambda$--term, and with a minimally coupled scalar field without
self--interaction.

For a spatially flat Friedman model
 \beq\label{21}
 ds^2=dt^2-a^2(t)(dx^2+dy^2+dz^2)
 \eeq
 and
 $$
 H=\frac{1}{a}\frac{da}{dt}\,{,}
 $$
 we get with $\zeta=\xi=0$
 $$
 a(t)\sim t^{1/3} \quad \mbox{and} \quad
 \phi =\phi_c\ln t
 $$
 where $\phi_c$ is an appropriately chosen constant. Thus, without the
correction terms, no inflationary solution can be found. However,
for $\zeta+4\xi>0$, an inflationary phase with
 \beq\label{22}
 \Lambda =\frac{1}{2(\zeta+4\xi)}
 \eeq
 exists, see \cite{CL}, Eq. (13). Let us now go into the details.
In \cite{CL} the metric (\ref{21}) was directly inserted into the Lagrangian (\ref{20}).
Here we first consider the field equation and insert the metric only
 afterwards. This has
the advantage that the details will become more clear.

\subsection{The Scalar Field Equation}

For the Lagrangian $L$, Eq. (\ref{20}), we get the field equation for $\phi$ as a
linear combination of $\Box \phi$ (from the usual metric term in Eq. (\ref{20})) and Eq.
(\ref{16}), where now $\alpha$ has to be replaced by $\xi/\zeta$, and (\ref{16}) has to
be multiplied by $\zeta$. One can directly see:
\begin{itemize}
  \item If $\phi_{,\, k}$ is covariantly constant and if $R$ is
  constant, then he equation is fulfilled.
  \item If $\phi_{,\, k}$ is covariantly constant and if
  $\alpha=-1/2$ (i.e. $\zeta=-2\xi$) then the equation is
  fulfilled.
\end{itemize}

Our main example is as follows: If $\phi=\phi_0t$, where $\phi_0$ is any constant and
$t$ is the time of metric (\ref{21}), then
 $$
 \phi_{;\, k}=(\phi_0, 0, 0, 0)\,{,} \quad \mbox{and}\quad
 \phi_{;\, kl}=-\Gamma^0_{kl}\phi_0\,{.}
 $$
 So that we get $\Box \phi=3H\phi_0$.

 {\bf Remark:} This implies that for $H\phi_0\neq 0$, the
 vector $(\phi_0,0,0,0)$ is not covariantly constant inspite
 of the constancy of $\phi_0$.

 Now we assume $H$ to be a positive constant, i.e., metric
(\ref{21}) represent the de Sitter space--time\footnote{To simplify the comparison, we
give here the known formalism for Einstein's theory: $\Lambda=3H^2$, $R_{ij}=-\Lambda
g_{ij}$, $R=-4\Lambda$, $R_{ij}-(R/2)g_{ij}=\Lambda g_{ij}$.}. To get a solution of the
scalar field equation, we have either
\begin{itemize}
  \item $\phi_0=0$;
\end{itemize}
or
\begin{itemize}
  \item $\phi_0\neq 0$ \quad \mbox{and} \quad
  $2\Lambda(\zeta+4\xi)=1$.
\end{itemize}
 The latter case represents Eq. (\ref{22}) above.

\subsection{ The Gravitational Field Equation}

Now we insert this de Sitter solution $g_{ij}$ and $\phi=\phi_0 t$ into the
gravitational field equation. The scalar $\psi$, Eq. (\ref{17}) now reads
 $$
 \psi=\phi_0^2=const
 $$
 From Eq. (\ref{18}) we get
 $$
 E_2^{ab}=4\Lambda \phi^{,\, a}\phi^{,\, b}-\Lambda g^{ab}\phi_0^2
 $$
 and a similar expression for $E_1^{ab}$.
 It turns out that for this highly symmetric case, the
gravitational field equation does not give an additional condition, so the de Sitter
space--time, with $\Lambda$ according to Eq. (\ref{22}), is a solution.

\subsection{The Stability of the de Sitter solution}

Let us now make the ansatz
 \beq\label{stab1}
 a(t)=e^{\alpha (t)}\,,
 \eeq
 where
 \beq\label{stab2}
 \alpha (t)=\alpha_0+H_0 t+\varepsilon\int \beta (t) dt\,, \quad
 \varepsilon\ll 1
 \eeq
 and
 \beq\label{stab3}
 \varphi (t)=\varphi_1+\varphi_0t+\varepsilon\int\gamma (t) dt\,,
 \eeq
 where
 \beq\label{stab4}
 \dot{\varphi}(t)=\varphi_0+\varepsilon\gamma (t)\,.
 \eeq
 Here $\varphi_0$, $\varphi_1$, $\alpha_0$, $H_0$ are constants,
the functions $\beta (t)$, $\gamma (t)$ have to be determined, and $\varepsilon$ is the
parameter of linearization.
 From Eqs. (\ref{stab1})--(\ref{stab4})  it follows that
 \beq\label{stab5}
 H=\frac{\dot{a}}{a}=H_0+\varepsilon\beta\,, \qquad
 \dot{H}=\varepsilon\dot{\beta}\,.
 \eeq
 and the Eqs. of motion [eq. (9)--(11) of the paper \cite{CL}] assume
the form
 \beq
 2\chi\varphi_0\varepsilon\ddot{\gamma}+4\eta H_0\varepsilon\varphi_0\dot{\gamma}+
 \varepsilon\varphi_0(1+6H_0^2\eta)\gamma+2\varepsilon
 (1+\eta\varphi_0^2)\dot{\beta}+3H_0^2(1+\eta\varphi_0^2)+\frac{\varphi_0^2}{2}=0
 \label{stab6}
 \eeq
 \beq
 6\chi\varepsilon\dot{\beta}-12H_0(\eta-3\chi)\varepsilon\beta+18\chi
 H_0^2-6\eta H_0^2+1=0 \label{stab7}
 \eeq
 \beq
 6\chi\varepsilon H_0\varphi_0\dot{\gamma}+
 \varphi_0\varepsilon(1+6H_0^2\eta)\gamma+
 6H_0\varepsilon
 (1+\eta\varphi_0^2)\beta+3H_0^2(1+\eta\varphi_0^2)+\frac{\varphi_0^2}{2}=0
 \label{stab8}
 \eeq
 where $\chi=-(2\xi+\zeta)$ and $\eta=-2(\xi+\zeta)$.
 The integration of (\ref{stab7}) and (\ref{stab8}) can be immediately
carried out leading to the solutions
 \beq\label{stab9}
 \beta(t)=\beta_0e^{c_1t}+c_2\,{,}
 \eeq
 where $\beta_0$ is a constant\footnote{From Eq. (\ref{23}) it will become clear that
in the region of parameters we are interested in both denominators of Eq. (\ref{stab10})
remain positive numbers.},
 \beq\label{stab10}
 c_1=-\frac{2H_0(4\xi+\zeta)}{2\xi+\zeta}\,,\quad c_2=
 \frac{1-6H_0^2(4\xi+\zeta)}{12H_0(4\xi+\zeta)},
 \eeq
 and
 \beq\label{stab11}
 \gamma(t)=-\frac{a_1a_3}{ba_1+aa_2}\, e^{c_1t}+
 \gamma_0e^{-(a_2/a_1)t}-\frac{a_4}{a_2}\,,
 \eeq
 where
 $$
 a=6\chi\varepsilon\,,\quad b=12H_0(\eta-3\chi)\varepsilon
 $$
 $$
 a_1=6H_0\varphi_0\chi\varepsilon\,,\quad
 a_2=(1+6H_0^2\eta)\varphi_0\varepsilon\,,\quad
 a_3=6(1+\eta\varphi_0^2)H_0\beta_0\varepsilon\,,
 $$
 $$
 a_4=6H_0(1+\eta\varphi_0^2)\varepsilon c_2+
 3(1+\eta\varphi_0^2)+\frac{\varphi_0^2}{2}
 $$
 We note that
 \beq\label{stab12}
 \frac{a_2}{a_1}=\frac{12H_0^2(\xi+\zeta)-1}{6H_0^2(2\xi+\zeta)}\,.
 \eeq
 Inserting (\ref{stab9}) and (\ref{stab11}) into (\ref{stab6}) one
gets a relation for these constants.
 As final comment, we infer the explicit expression of
 \beq\label{stab13}
 \alpha
 (t)=(\alpha_0+\varepsilon C_0)+(H_0+c_2)t+\varepsilon\beta_0c_1e^{c_1t}
 \eeq
 and
 \beq\label{stab14}
 \varphi(t)=\varphi_1+\left(\varphi_0-\varepsilon\frac{a_4}{a_2}\right)t-
 \varepsilon\frac{a_1a_3}{(ba_1+aa_2)c_1}e^{c_1t}-\varepsilon\gamma_0\frac{a_1}{a_2}
 e^{-(a_2/a_1)t}\,{.}
 \eeq
 The main point of the deduction is that $c_1$ from Eq. (\ref{stab10}) is a negative
real number.

 Immediately we see that the conditions for the stability of the de
Sitter are satisfied
 \beq\label{stab15}
 \frac{\dot{H}}{H^2}\to 0\,,\quad \frac{\varphi(t)}{t}\to
 0\,,\quad \mbox{as}\quad t\to \infty\,.
 \eeq

\section{Conclusions}
\setcounter{equation}{0}

Which values of $\zeta$ and $\xi$ in eq. (\ref{20}) will give sensible results? As it is
always the case in such higher--derivative theories, several ranges of the parameters
have to be excluded. We have already seen that for $\zeta=-2\xi$ we have a singular
point of the differential equation and that we need $4\xi+\zeta>0$ to ensure
$\Lambda>0$. So it seems to be adequate that we require
 $\zeta>-4\xi$ and $\zeta\neq -2\xi$.
However, we will require a little stricter
\beq\label{23}
\zeta > 4\vert\xi\vert \geq 0\,.
\eeq

 To discuss the stability of the de Sitter solution, one can
compare it with other Friedman solutions; this has already been done in \cite{CL}, where
the de Sitter space--time has found to be a solution. (For ease of comparison: Eq.
(\ref{23}) implies $A<0$ and $B>0$ in \cite{CL}, Eq. (16).) Then the field equation has
solutions
 $$
 H=c_1\tanh (c_2 t) \quad \mbox{and} \quad H=c_1\coth (c_2 t)\,{,}
 $$
 ($c_1, c_2>0$) both having $H\to const >0$ as $t \to \infty$.

 A more thorough discussion of the stability can be performed if
the symmetry of the metric is not prescribed from the beginning. In order that the
Cauchy problem be well--posed, we need some further conditions. (However, these
conditions are fulfilled in a neighborhood of the interesting de Sitter space--time.)

 The scalar field equation is of second order in $\phi$, so it
has the structure
 \beq\label{24}
 \phi_{,\, 00}=F(\phi, \phi_{,\, 0}, R_{ij}, R_{,\, 0})
 \eeq
 (The dependency on $R_{,\, 0}$ disappears for
$\alpha=-1/2$, see Eq. (\ref{16}), but this case is not covered by the allowed cases Eq.
(\ref{23}), and the dependency on the spatial derivative is not explicitly mentioned.)
This equation (\ref{24}) can be derivated with $d/dt$, and one gets
 \beq\label{25}
 \phi_{,\, 000}=G_1(\phi, \phi_{,\,0}, R_{ij}, R_{ij,
 \,0}, R_{,\, 0}, R_{,\, 00})
 \eeq
 The right--hand sides of Eqs. (\ref{24}), (\ref{25}) have to be
inserted into the gravitational field equation to replace the artificial second and
third derivative of $\phi$. Thus a fourth--order field equation for $g_{ij}$ results
with leading--order term $R_{,\, 00}$.

 A more detailed elaboration of the corresponding stability has to
be done yet. Further, let us mention that the terms $L_1$ and
$L_2$ discussed above, can be related to the trace anomaly.

Finally, there seems strong evidence that the model discussed in
this paper is not related by any conformal transformation to any
known model. A further step for finding related models might be to
rewrite the metric in a more general transformation of the form
 \beq\label{26}
 \tilde{g}_{ab}=g_{ab}+\lambda Rg_{ab}+\mu R_{ab}
 \eeq
 with $\lambda$ and $\mu$ constants. (The Einstein tensor for
$\tilde{g}_{ab}$ gives a tensor of order 4 if rewritten with $g_{ab}$).
Another idea
goes as follows: Following \cite{Sok} (see the Introduction), we write (Eq. (\ref{20})
 \beq\label{27}
 \tilde{g}^{ab}=\frac{\partial L}{\partial
 R_{ab}}=-\frac{1}{2}g^{ab}+\zeta \phi_{,\, a}\phi_{,\, b}+\xi
 g^{ab} \phi_{,\, k}\phi^{,\, k}
 \eeq
 which gives a transformation also of a more general structure: It
combines a conformal transformation with a Schild--transformation
of type Eq. (\ref{5}).
 It is not clear at the moment whether one of these
transformations will simplify the equation or not. Probably we need an additional tensor
field instead of an additional scalar field to be able to transform to Einstein's
theory. The appearance of an inflationary solution in the nonminimal derivative coupling
model discussed here can be explained as follows: In regions, where
 $$
 R_{ij}\sim -\Lambda g_{ij} \qquad (\Lambda >0)
 $$
 and
 $$
 \phi_{,\, k} \sim (\phi_0, 0, 0, 0), \quad
 (\phi_0\neq 0)
 $$
 the scalars $R^{ij}\phi_{, \, i}\phi_{, \, j}$ and $R\phi_{,\, k}
\phi^{,\, k}$ both are negative and approximately constant. So, in this approximation,
their appearance in the Lagrangian mimics an effective cosmological constant.

\bigskip
\bigskip

\centerline{\bf Acknowledgement}

HJS thanks the colleagues of Salerno University where this work has been done,
especially  Gaetano Scarpetta for kind hospitality and the pleasant atmosphere at the
Dept. of Science Fisiche E.R. Caianiello.

\end{document}